# Comment on "Microwave vortex dissipation of superconducting Nd-Ce-Cu-O epitaxial films in high magnetic fields"

Pieder Beeli

Colorado State University; Dept. of Physics; Fort Collins, CO 80523

pbeeli@lamar.colostate.edu

**Abstract**

While the electrodynamic treatment of a superconducting thin film by Yeh *et al.* [Phys. Rev. B **48**, 9861 (1993)] ostensibly has application for the "*thin-film* limit," their work is shown to employ *bulk-like* instead of *thin-film* electrodynamics, and as such obtains an ostensible violation of the first law of thermodynamics and a surface resistance independent of the resistivity. Other theoretical and experimental difficulties are briefly noted.

## I. INTRODUCTION

Ref. [1] concerns the thin-film limit of electrodynamic dissipation in a superconductor. It's theory alleges to have application for films with "a *general thickness*" and all the fits employ a thin-film limit of this formalism. This comment focuses on a central error in [1]: *Bulk-like* electrodynamics are used in lieu of *thin-film* electrodynamics. Because this contradiction is foundational to [1], we limit our focus to this and make only passing reference to other difficulties. Although there has been an erratum[2] published associated with [1], the difficulties noted herein are not addressed by the erratum. Indeed the erratum claims that, "all the physics and analyses presented in the paper remains [sic] unchanged."

## II. CORRECT AND INCORRECT ELECTRODYNAMIC EXPRESSIONS

We follow the convention in [1] which uses a spacial-temporal representation of $\exp(z/\tilde{\lambda} - i\omega t)$, takes the *E*-field to be in the *y*-direction, the *H*-field to be in the *x*-direction and aligns the side of the film which faces the normally incident radiation at $z = d$ (where *d* is the film thickness). To understand the electrodynamic development of [1], we digress to study bulk fields and impedance, next, the fields for films of arbitrary thickness and then the impedances for arbitrary thicknesses.

## II. a. Bulk Limit Fields and Impedance

In the bulk limit, the *H*-field expression given in [1]: $H_x(z) = h_f \exp(z/\tilde{\lambda} - i\omega t)$ is applicable. From Lenz's Law, $\nabla \times \dot{E} = -\hat{x} E_y / \tilde{\lambda} = \hat{x} i \omega \mu H_x$, it therefore follows that the bulk limit surface wave impedance $Z_{swb}(z = d)$ is

$$Z_{swb}(z = d) \equiv \frac{E_{yb}(z = d)}{H_{xb}(z = d)} = -i\mu_o \omega \tilde{\lambda}. \quad (1a)$$

From Ampere's Law and by making the good conductor approximation we obtain $\nabla \times \hat{H} = \hat{y} H_x / \tilde{\lambda} \approx \hat{y} \tilde{\sigma} E_y$. From this we obtain a near equivalent expression

$$Z_{swb} \equiv \frac{E_{yb}(z = d)}{H_{xb}(z = d)} \approx \frac{1}{\tilde{\sigma} \tilde{\lambda}} \quad (1b)$$

where

$$Z_{swb} = R_{swb} - i X_{swb} \quad (1c)$$

where $R_{swb}(X_{swb})$ is purely real and is the bulk surface wave resistance(reactance) at the incident surface.

## II. b. Correct and Incorrect Form of the Fields for Arbitrary Film Thicknesses

Since both *E* and *H* obey a Helmholtz equation, it follows that inside the film the phasor *H*-field *should* have the form:[3]

$$H_x(z) = (h_f e^{-(d-z)/\tilde{\lambda}} + h_r e^{(d-z)/\tilde{\lambda}}) e^{-i\omega t} \quad (2a)$$

for $d \leq z \leq 0$ where $h_{f(r)}$ is the steady-state complex forward(refluent) *H*-field amplitude (where $h_f + h_r = h_o$ where $h_o$ is the total complex tangential *H*-field amplitude at the incident surface of the superconductor). $\tilde{\lambda}$ is the complex penetration depth. $z = d$ marks the incident surface of the Nd-Ce-Cu-O film and $z = 0$ marks the interface between the Nd-Ce-Cu-O film and a $LaAlO_3$ substrate. By employing Ampere's Law and making the good conductor approximation, $\nabla \times \tilde{H} = \hat{y} \, \partial H_x/\partial z \approx \hat{y} \tilde{\sigma} E_y$. From this and Eq. 1, it is seen that the phasor *E*-field *should* have the form:[3]

$$E_y(z) \approx -i\mu_o \omega \tilde{\lambda}(h_f e^{-(d-z)/\tilde{\lambda}} - h_r e^{(d-z)/\tilde{\lambda}}) e^{-i\omega t} \quad (2b)$$

for $d \leq z \leq 0$. In contrast to Eq. 2b, Eq. 3 of [1] is:

$$E_y(z) = -i\mu_o \omega \tilde{\lambda} h_o e^{-i\omega t} e^{-(d-z)/\tilde{\lambda}} \quad (3)$$

for $d \leq z \leq 0$. Eq. 3 is a bulk electrodynamic equation. It's bulk character, as we shall discuss next, is manifest in both the implicit expression for the surface wave impedance and in the omission of the refluent field.

## II. c. Correct and Incorrect Form of the Surface Wave Impedance

Next we seek to compare both the bulk $Z_{sw}$ and the arbitrary thickness $Z_{sw}$ to a $Z_{sw}$ result from [1]. But before we compare these quantities, we must first determine them.

### II. c. 1. Surface Wave Impedance Expression from Ref. 1

By comparing Eq. 3 to Eq. 2, we see that the $Z_{sw}$ implicit in Eq. 3 is:

$$Z_{sw} = -i\mu_o \omega \tilde{\lambda} . \quad (4)$$

As the development in [1] ostensibly applies for "a *general thickness*," it therefore claims that Eq. 4 is the surface wave impedance for *all* film thicknesses.

### II. c. 2. Bulk Surface Impedance Expression

An examination of Eq. 4 reveals that it is just the surface wave impedance *if* the superconducting film was in the *bulk* limit (*i.e.* Eq. 1a). Since negligible errors are introduced by making the good conductor approximation in a *bulk* effective medium superconductor,[3,4] *were* the superconducting film in the *bulk* limit, it *would* be true that

$$Z_s \equiv \frac{E_y(z=d)}{K_y(z=d)} \approx -i\mu_o \omega \tilde{\lambda} \quad (5a)$$

where *K* is the surface current density.[3] Further,

$$Z_s = R_s - iX_s \quad (5b)$$

where $R_s(X_s)$ is purely real and is the surface resistance(reactance). However, Eqs. 4, 5a and 5b are *bulk* equations and as such yield values for $Z_{sw}$ and $Z_s$ that are *grossly different* than the thin-film equations applicable to [1].

### II. c. 3. Correct Surface Wave Impedance Expression for the Work of Ref. 1

From Eq. 2 it is clear that the correct arbitrary thickness expression for $Z_{sw}$ is

$$Z_{sw} \equiv \frac{E(z=d)}{H(z=d)} \approx -i\mu_o \omega \tilde{\lambda} \left(\frac{h_f - h_r}{h_f + h_r}\right) \quad (6)$$

which is quite different than Eq. 4--the result of [1].

## III. POWER DISSIPATION

### III. a. Correct Thin-Film Power Dissipation Expressions

Next we turn to the more general thin-film electrodynamics to determine the steady-state power dissipation of TEM radiation. Ref. 1 uses the Coffey-Clem electrodynamic model for large magnetic fields (Eq. 5 of ref. 1) and the Ambegaokar-Baratoff model (Eq. 8 of ref. 1) for small fields. The power dissipated in a film ($P_{film}$) with area (*A*) and thickness (*d*) can be found by[3,4]

$$2P_{film} = A \int_d^0 \Re(J_{film} \cdot E_{film}^*) dz . \quad (7)$$

(Although Eq. 7 is used in [1] in conjunction with the Coffey-Clem model, the *E*-field is given by Eq. 3-- which we have shown to be incorrect.)

Alternatively, consider the incident power ($P_{in}$) to be emerging from free space (for simplicity) onto a structure composed of many layers. *Thin-film* electrodynamics indicates that for an incident power level of $P_{in}$, the steady-state power dissipating in the $k^{th}$ layer ($P_k$) of an *n*-layered system is[5]

$$P_k/P_{in} = [|\Psi_{ki}|^2 \Re(Z_{ki}) - |\Psi_{kb}|^2 \Re(Z_{kb})]/\eta_o \quad (8)$$

where $1 \leq k < n$ and $\Psi_{ki(b)}$ is the ratio of the complex phasor *H*-field amplitude at the incident(back) surface of the $k^{th}$ film to the free-space incident *H*-field. $\Re(Z_{ki[b]})$ is the real part of the surface wave impedance at the incident[back] surface of the $k^{th}$ layer. Eq. 8 expresses the power dissipating in the film as a difference between the power crossing the incident surface of the film and the power crossing the back surface of the film.

Taking the limit of Eq. 8 in which the fields at the end of the $k^{th}$ layer are negligible, we obtain a bulk expression:

$$P_k = A|H_k|^2 R_{ksw} \quad (9a)$$

where $|H_k|$ is the peak magnitude of the amplitude of the tangential *H*-field on the incident surface of the $k^{th}$ layer and $R_{ksw}$ is the real part of the surface wave impedance at the incident surface of the $k^{th}$ layer.[6] In the limit that the $k^{th}$ layer goes bulk *and* satisfies the good conductor approximation, the surface wave impedance is well-approximated by the surface impedance.[3] Under these conditions, $P_k$ is also given by:

$$P_k \approx A|H_k|^2 R_{ks} \qquad (9b)$$

where $R_{ks}$ is the surface resistance at the incident surface of the $k^{th}$ layer.

## III. b. The Bulk-like Power Dissipation Expression used in Ref. 1

There are three contradictions associated with a bulk power dissipation expression from ref. 1. The first contradiction concerns the *defining* of a thin-film system to be bulk. The next two contradictions arise when interpreting Eq. 10 either in terms of Eq. 9b or Eq. 9a.

## III. b. 1. The Thin-Film System of Ref. 1 is *Defined* to be Bulk

Ref. 1 incurs a contradiction by invoking a power dissipation expression for a thin-film application. With the low-field treatment involving the Ambegaokar-Baratoff model, instead of using Eq. 8, ref. 1 uses something resembling Eq. 9b:

$$R_s \equiv P_{tot}/(Ah_o^2). \qquad (10)$$

Similar to the mistake noted in sec. II. c. 2 where bulk-like *field* expressions are used instead of thin-film *field* expressions (*i.e.*, Eq. 3 instead of Eq. 2), [1] incorrectly *defines* the *power* dissipation to be governed by a bulk expression instead of by a thin-film expression (*i.e.,* by Eq. 10 instead of Eq. 7 or Eq. 8).

Since [1] relates the power dissipation to $R_s$ via Eq. 10, the mistake of employing this bulk power dissipation expression leads to a mistake in the surface resistance expression. In this vein, Eq. 10 yields a normal state $R_s$ to be:

$$R_s = \mu_o \omega d \qquad (11)$$

where $\omega$ is the angular frequency, $\mu_o$ is the permeability of free space and $d$ is the thickness of the superconducting film. Eq. 11 is discussed at length in sec. IV. e.

## III. b. 2. The use of Surface Resistance when the Good-Conductor Approximation is not Satisfied

Ref. 1 incurs a contradiction by invoking $R_s$ in Eq. 10--interpreted in the sense of Eq. 9b--because the good-conductor approximation is not satisfied for regions where appreciable power is dissipating. That Eq. 10 *cannot* be understood in terms of Eq. 9b is immediately seen from the thin-film limit of Eq. 3 which indicates that the field amplitude incident on the LaAlO$_3$ substrate (which does not satisfy the good-conductor approximation) are comparable to those incident on the Nd-Ce-Cu-O thin-film. From this claim, it follows that $H_x(z = d)$ cannot be approximated by $K_y(z = d)$; therefore, Eq. 9b contradicts the thin-film limit of Eq. 3.

## III. b. 3. The Power in the Superconducting Thin-Film is Equated to the Power Dissipating in a Bulk System.

But ref. 1 also incurs a contradiction by interpreting Eq. 10 in the sense of Eq. 9a. Interpreted in this manner, Eq. 10 would represent the power dissipating in many layers including the Nd-Ce-Cu-O thin film, the LaAlO$_3$ substrate and whatever other unspecified backing materials are present in the experiment of [1]. This results in a contradiction because ref. 1 equates the dissipation in the film only (*i.e.*, Eq. 7) to the dissipation over many materials (*i.e.*, Eq. 10 or Eq. 9a).

## III. c. Ostensible Violation of the First Law of Thermodynamics

Ref. 1 predicts a violation of the first law of thermodynamics. To see this prediction one can use either of two (contradictory) $R_s$ expressions given in [1] (*i.e.*, either Eq. 4 or Eq. 11).

This violation is a product of the claims made by [1] regarding the fields (*i.e.*, Eq. 3) and the impedance (*i.e.*, Eq. 4 or Eq. 11). Having established that according to the thin-film limit of Eq. 3, the fields on either side of the superconducting thin film are comparable, $|\psi_{ki}| \approx |\psi_{kb}|$. From Eq. 4 *or* Eq. 11 it follows that $R_s = \Re(Z_{swi}) \ll \Re(Z_{swb})$ where $Z_{swb}$ is the surface wave impedance (in the forward direction) at the *back* of the Nd-Ce-Cu-O film (or equivalently, at the incident surface of the LaAlO$_3$ substrate). Therefore, from Eq. 8 it follows that [1] predicts *negative* power dissipates in the superconducting thin-film or equivalently that the superconductor does not *dissipate* net energy, but *produces* it (in violation of the first law of thermodynamics).

## IV. THE FIELD OR IMPEDANCE EXPRESSIONS OF REF. 1

Sec. II. c. 2 shows that a $Z_{sw}$ result from [1] (*i.e.,* Eq. 4) is identical to the bulk result. Since [1] claims to perform the mathematical analysis for films of "a *general thickness*," we conclude that the mathematical modeling of [1] is incorrect for thicknesses other than the bulk limit. Since Eq. 4 is a bulk expression, in sec. IV we consider the validity of approximating Eq. 4 with the correct expression, $Z_{sw}$ of Eq. 6.

But first, to gain a heuristic appreciation for the thin film limit especially vis-a-vis $Z_{sw}$ and $Z_s$, we shall consider a *gedanken* experiment where the thickness of the film becomes vanishingly thin.

## IV. a. *Gedanken* Experiment for $Z_{sw}$ and $Z_s$ in the limit $d \rightarrow 0$

Since [1] alleges to derive the electrodynamic expressions for "a *general thickness*," it ought to recover the correct result in the limit that $d \rightarrow 0$. That the correct expressions of $Z_{sw}$ and $Z_s$ are far different than $-i\mu\omega\tilde{\lambda}$ (Eq. 4) can also be intuitively seen by a *gedanken* experiment where we imagine the superconducting thin film to be suspended in free space and consider the limit that $d \rightarrow 0$. In this limit it must be true that $Z_{sw} = \eta_o$ (where $\eta_o$ is the intrinsic impedance of free space and is equal to about 377 Ω) and that $Z_s = \infty$ (since there is no conduction current). Just as $\eta_o$ and $\infty$ are grossly different from the value given by Eq. 4 in this *gedanken* experiment, so are the correct values of $Z_{sw}$ and $Z_s$ very different from Eq. 4 in the thin film system of [1]. But if the thin film $Z_{sw}$ and $Z_s$ are very different from that of Eq. 4, then it must be the case that Eq. 3 is not even approximately valid. Although $Z_{sw}$ is correctly given by Eq. 6, $h_f$ and $h_r$ are *indeterminable* without knowledge of the material properties for the region $z < 0$. The correct expression for $Z_{sw}$ could be obtained by a transmission line analog as done in [3, 4, 5 and 7], but *cannot be obtained from [1]* because the thickness of the LaAlO$_3$ substrate, *inter alia,* is not given.

## IV. b. Bulk Character of $Z_{sw}$ Determined from the Thin-Film Limit of Eq. 3

The remainder of sec. IV discusses contradictions between $Z_{sw}$ *per the mathematical analysis of [1]* and the thin-film system that is studied *per the text of [1]*.

### IV. b. 1. Bulk-like $Z_{sw}$ in a Thin-Film Limit

This usage of a *bulk* form of $Z_{sw}$ (*i.e.*, Eq. 4) in the *thin-film* system of [1] is a contradiction. Per the discussion of sec. II, *Eq. 3 claims that the Nd-Ce-Cu-O film is in the bulk limit for all film thicknesses including the thin-film limit*. Since as discussed in sec. IV. a the thin-film limit of Eq. 3 must possess a thin-film expression for $Z_{sw}$, this is a contradiction.

### IV. b. 2. The Good-Conductor Approximation

As discussed in sec. III. b. 2, Eq. 3 also possesses a contradiction in light of the good conductor approximation. This contradiction of Eq. 3 emerges in the thin-film limit which is explicitly considered in *all* the $R_s$ fits of [1]. In this limit Eq. 3 reveals that the fields incident on the LaAlO$_3$ substrate--where the good-conductor approximation is not satisfied--are *comparable* to the fields incident on the Nd-Ce-Cu-O film. Therefore, $H_x(z=d)$ cannot be approximated by $K_y(z=d)$ [3] so that $Z_{sw}$ (*i.e.,* Eq. 4) *cannot* be approximated by $Z_s$ (Eq. 5a) and therefore *neither $Z_{sw}$ nor $Z_s$ can be approximated by* $-i\mu\omega\tilde{\lambda}$ [8]. If the fields at the back of the film are comparable in magnitude to the fields at the front then the film cannot be in the thick-film limit and must therefore have an associated thin-film expression for $Z_{sw}$. Thus while Eq. 3 implies Eq. 4, Eq. 4 cannot be correct and therefore neither can Eq. 3 be correct. Thus the thin-film limit of Eq. 3 simultaneously claims that Eq. 4 is both applicable (by virtue of the prefactor of Eq. 3) and inapplicable (since *neither $Z_{sw}$ nor $Z_s$ can be approximated by* $-i\mu\omega\tilde{\lambda}$ [8]), or that the film is in *both* the bulk *and* thin-film limit.

### IV. b. 3. The Irrelevance of Underlying Materials in a Thin-Film System

Ref. 1 incurs a contradiction by purporting to do quantitative thin-film electrodynamics while failing to even provide a quantitative electrodynamic *description* of the other materials in the thin-film structure studied. This is especially unfortunate because as noted earlier, Eq. 3 claims that the fields penetrating into the LaAlO$_3$ substrate are comparable to the fields on the incident surface of the Nd-Ce-Cu-O film. This begs the question: *How is it that although most of the microwave energy is transmitted through the film, the material parameters of the dominant dissipative media are irrelevant to the quantitative analysis of [1]?*

As the film becomes increasingly thin, the backing materials become increasingly important and must bear on the total power dissipation and on the dissipation in the superconducting thin film.[3,4] Recent work on bimetallic structures reveal numerous limiting forms of $Z_s$ which depend on both the impedance mismatch between the metals and the thickness of the film.[9] Expressions for $Z_s$ for both bimetallic structures[3] and superconducting film-bulk metal structures[4] reveal that in the limit that $d \to 0$, $R_s$ will be that of the backing metallic structure alone. Unfortunately neither Eq. 11 nor the $Z_{sw}$ inferred from Eq. 3 (*i.e.*, Eq. 4) manifest this property even though dielectric substrates can exhibit considerable dissipation.[5,10] Microwave experiments have revealed pronounced effects on the total dissipation due to a substrate underlying a superconducting thin film;[7] thus we expect [1] to manifest *some* effect of the LaAlO$_3$ substrate, especially since the case "$d \ll \lambda_{eff}$" is explicitly considered.

## IV. c. The Composition of the Thin-Film Structure at $z = 0^-$

There is a contradiction in the characterization of the *in-situ* thin-film structure as the absence of a refluent field in Eq. 3 means that the spatial form of Eq. 3 is bulk-like and not thin-film like. Per the *text* of [1], at $z = 0$ there is an interface between the Nd-Ce-Cu-O film and the LaAlO$_3$ substrate. However the representation of the *H*-field by a single (forward) term in Eq. 3, *only* applies for a plane wave incident on one side of a *bulk* structure. Eq. 3 therefore indicates that instead of there being a *substrate* at $z = 0^-$ (which locates the LaAlO$_3$ side of the interface with the superconducting thin film), there is actually a *superconductor* there with the identical $\tilde{\lambda}$ as the Nd-Ce-Cu-O film and with a thickness thick enough so that the fields have attenuated to a negligible amplitude on the side opposite to the incident side. Eq. 3 represents a bulk superconductor with a thickness greater than the amplitude (attenuation) length scale ($\delta_A$), where $\delta_A \equiv |\tilde{\lambda}|^2 / \Re(\tilde{\lambda})$ where $\Re$ is the real operator [4]. Eq. 3 indicates that instead of a LaAlO$_3$ dielectric occupying the space from $z = 0$ to some unspecified location (as per the *text* of [1]), there is actually a superconductor extending from $z = 0$ to some value of $z$ which is at least $\sim 3\delta_A$ less than $z = 0$. (In the notation of [1], $\delta_A = \lambda_{eff}$.)

## IV. d. Reducing Thin-Film Electrodynamics to a Subset of Bulk Electrodynamics

Ref. 1 incurs an epistemological contradiction in describing thin-film electrodynamics as a subset of bulk electrodynamics, instead of vice-versa. Because [1] *begins* with bulk-like electrodynamics in describing the fields, the ensuing attempt to derive thin-film expressions for $R_s$ and the power dissipation is destined to be incorrect. By examining Eq. 2, one can see that in the bulk limit $h_r$ must be identically zero. Were this not the case, the energy density would be infinite. Therefore the bulk limit is characterized by the presence of only negligible fields at the far side of the sample. This discussion makes it clear that *bulk* electrodynamics is a special case of *thin-film* electrodynamics and that [1] has instead attempted to make *thin-film* electrodynamics a special case of *bulk* electrodynamics.

## IV. e. Expression for the Surface Resistance

As briefly noted in sec. III. c, ref. 1 advocates two contradictory expressions for the normal state surface resistance in the thin-film limit. One is given by Eq. 4 and the other by Eq. 11. The former is the *bulk* result and the latter is unphysical in any limit. The erratum[2] only reaffirmed the commitment of Yeh *et al.* to Eq. 11 as these authors rescaled the thin-film data so as to be consistent with Eq. 11. Being independent of the resistivity, Eq. 11 is clearly wrong, claiming that all superconducting films in the normal state and which have the same thickness, have the same $R_s$--even if their resistivities differ by orders of magnitude. (The normal state limit of $R_s$ of a bulk superconductor can be expressed as $R_s \approx \mu_o \omega \delta / 2$ where $\delta$ is the normal state skin depth and is given by $\delta = \sqrt{2\rho / \mu_o \omega}$, which is proportional to $\sqrt{\rho}$ where $\rho$ is the normal state resistivity.) Apart from the $\rho$-independent character of Eq. 11, Eq. 11 contradicts the result of Eqs. 4 and 5, which holds that $R_s \approx \mu_o \omega \delta / 2$ in the normal state.

## IV. f. Omission of any Boundary Conditions

The electrodynamic fields in [1] are transverse. From Maxwell's curl equations it follows that *E* and *H* must be continuous. Yet we note that the electrodynamic development in [1] does not satisfy a single boundary condition. This omission of boundary considerations is further support of the thesis that [1] employs bulk instead of thin-film electrodynamics

## V. OTHER DIFFICULTIES WITH REF. 1

Having shown in detail that [1] erroneously employs *bulk* electrodynamic concepts to understand *thin-film* electrodynamics and therefore has it's foundational claim of thin-film electrodynamics invalidated, other difficulties are briefly noted. Although [1] provides citations for a pinning force constant, it uses a field-*dependent* expression when the citations employ a field-*independent* expression. No justification for the insertion of this field dependence is given.

Experimentally two difficulties with [1] are mentioned. It is known that the zero-field resistivity vs. temperature transition is the same for dc or microwave measurements on a high-quality cuprate crystal,[11] but yet Fig. 1 of [1] shows $T_c$ being *lowered by several degrees* at microwave frequencies. This may be due to operating the resonator at a too high power level such that large current densities are induced and suppress the transition. The second experimental difficulty concerns the *higher* frequency $R_s$ data of Fig. 2b which have a *lower* value--for B > 0 T and T < 10 K--than the *lower* frequency data of Fig. 2a. This observation contradicts what one would expect from Lorentz force dissipation and Eq. 3, which is that $R_s$ should *increase* with frequency.

## VI. CONCLUSION

Because [1] uses bulk expressions which are integrated over a material with an arbitrary thickness $d$, its treatment is *neither* applicable for a bulk material *nor* a thin film.

[1] incurs four contradictions involving the power dissipation: it uses a bulk-like expression for the power dissipation (Eq. 10) for a thin-film system, uses an inappropriate power dissipation expression involving $R_s$ (Eq. 10) when the good conductor approximation is not satisfied over regions where substantial power is dissipating and equates the power dissipating in the superconducting thin film with the power dissipation of a bulk structure. Finally an exact electrodynamic thin-film expression (*i.e.*, Eq. 8) is used to show that [1] predicts a violation of the first law of thermodynamics.

Atop of these four issues involving the power dissipation, there lies more involving $Z_{sw}$ and the fields. It has been shown that $Z_{sw}$ of [1] contradicts intuitive limits of a *gedanken* experiment and that at least three contradictions result from the bulk character of $Z_{sw}$ found from the thin-film limit of Eq. 3: the *bulk* character of $Z_{sw}$ contradicts the *thin-film* result for $Z_{sw}$, the presence of substantial fields at the surface of the LaAlO$_3$ contradicts the good-conductor-approximation character of $Z_{sw}$ and there is the contradiction of the thin-film limit of Eq. 3 which, having a bulk form of $Z_{sw}$, suggests that underlying materials are irrelevant. This contradictory notion of the irrelevance of underlying materials is twice more reaffirmed: in omitted any quantitative description of these underlying materials and in failing to consider a single boundary condition. Eq. 3 contradicts the *rhetorical* claim of [1] that $z = 0^-$ marks the location of LaAlO$_3$. [1] incurs an epistemological contradiction in that it expresses thin-film electrodynamics as a subset of bulk electrodynamics (instead of vice-versa). [1] provides contradictory expressions for the normal state thin-film limit of $Z_{sw}$: $\mu_o\omega\delta/2$ and $\mu_o\omega d$. Finally this latter result, being independent of the resistivity, is clearly unphysical.

Unfortunately the problems for [1] do not stop here. We have noted a ~5 K suppression of $T_c$ for the microwaves from the $T_c$ found from dc resistivity data. This suggests that the microwave power may have exceeded the threshold for the linear regime. The *higher* frequency $R_s$ data of Fig. 2b of [1] have a *lower* value--for B > 0 T and T < 10 K--than the *lower* frequency data of Fig. 2a of [1]. This observation contradicts what one would expect from Lorentz force dissipation and Eq. 3, which is that $R_s$ should *increase* with frequency.

It is seen that the field expressions of [1] are inapplicable for thin-film electrodynamics, and that these field expressions predict at least a dozen contradictions or unphysical results including a violation of conservation of energy. Having failed to account for the geometric aspects of this problem, all the fitting parameters (*e.g.* critical fields, force pinning constant, pinning potential and London penetration depth) claimed by [1] are of dubious value.

## VII. ACKNOWLEDGEMENTS

The author thanks R. O. Waters and C. Mathieu for proofreading, N.-C. Yeh, D. M. Strayer and R. P. Vasquez for noting a sign error in an earlier version of this manuscript and H. A. Blackstead for bringing his attention to [1]. This work was performed at the University of Notre Dame.

## VIII. ENDNOTES

sion in [1] about this changing measure for the field amplitudes.